%% file: main.tex
\documentclass[acmsmall,screen]{acmart}  
\AtBeginDocument{%
  }

\setcopyright{acmlicensed}
\copyrightyear{2018}
\acmYear{2018}
\acmDOI{XXXXXXX.XXXXXXX}
\acmConference[Conference acronym 'XX]{Make sure to enter the correct
  conference title from your rights confirmation email}{June 03--05,
  2018}{Woodstock, NY}
\acmISBN{978-1-4503-XXXX-X/2018/06}

\usepackage[utf8]{inputenc}
\usepackage{newunicodechar}
\newunicodechar{：}{:}

\usepackage[table]{xcolor} 
\usepackage{booktabs}
\usepackage{graphicx}
\usepackage{multirow}
\usepackage{makecell}
\usepackage{graphicx}
\usepackage{subcaption}
\usepackage{graphicx}
\usepackage{caption}
\usepackage{tcolorbox}
\usepackage{xcolor}
\usepackage{wrapfig,lipsum,booktabs}
\tcbuselibrary{skins} 
\usepackage{xcolor}
\usepackage{seqsplit}
\usepackage{amsmath} 

\newtcolorbox{summarybox}[1][]{
    enhanced,            
    sharp corners,
    boxrule=0pt,
    frame hidden,       
    borderline west={3pt}{0pt}{black},
    colback=gray!10,
    left=3pt,
    right=3pt,
    top=3pt,
    bottom=3pt,
    #1
}

\newcommand{\smell}{\textit{Interaction Smells}}
\newcommand{\tool}{\textit{InCE}}
\begin{document}


\title{An Empirical Study of Interaction Smells in Multi-Turn Human-LLM Collaborative Code Generation}



\author{Binquan Zhang, Li Zhang, Lin Shi, Song Wang, Yuwei Qian, Linhui Zhao, Fang Liu, An Fu, Yida Ye}

\renewcommand{\shortauthors}{Trovato et al.}


\include{sec/0_abs}

\ccsdesc[500]{Software and its engineering~Software development techniques}




\maketitle

\input{sec/1_intro}

\input{sec/2_bg}

\input{sec/3_method}
\input{sec/4_smells}
\input{sec/5_test_LLM}

\input{sec/6_InCE}
\input{sec/7_discussion}

\input{sec/8_conclusion}


\bibliographystyle{ACM-Reference-Format}
\bibliography{my_ref}

\end{document}

%% file: sec/0_abs.tex
\begin{abstract}
Large Language Models (LLMs) have revolutionized code generation, evolving from static tools into dynamic conversational interfaces that facilitate complex, multi-turn collaborative programming. 
While LLMs exhibit remarkable proficiency in generating standalone code snippets, they often struggle to maintain contextual consistency during extended interactions, creating significant obstacles in the collaboration process. 
Existing benchmarks primarily emphasize the functional correctness of the final output, overlooking latent quality issues within the interaction process itself, which we term {\smell}. 
In this paper, we conduct an empirical study on sampled real-word user-LLM interactions from WildChat and LMSYS-Chat-1M datasets to systematically investigate {\smell} in human-LLM code generation tasks from the perspectives of phenomena, distribution, and mitigation. 
First, we establish the first taxonomy of {\smell} by manually performing open card sorting on real-world interaction logs. 
This taxonomy categorizes {\smell} into three primary categories, i.e., \textit{User Intent Quality}, \textit{Historical Instruction Compliance}, and \textit{Historical Response Violation}, comprising nine specific subcategories. 
Next, we quantitatively evaluate six mainstream LLMs (i.e., GPT-4o, DeepSeek-Chat, Gemini 2.5, Qwen2.5-32B, Qwen2.5-72B, and Qwen3-235B-a22b) to analyze the distribution of {\smell} across different models. 
Finally, we propose Invariant-aware Constraint Evolution (InCE), a multi-agent framework designed to improve multi-turn interaction quality through explicit extraction of global invariants and pre-generation quality audits. 
Experimental results on the extended WildBench benchmark demonstrate that this lightweight mitigation approach significantly improves the Task Success Rate and effectively suppresses the occurrence of {\smell}.


\end{abstract}

%% file: sec/1_intro.tex
\section{Introduction}
With the rapid advancements in Large Language Models (LLMs), code generation has evolved beyond the mere translation of natural language descriptions into isolated snippets. Instead, it has evolved into human-LLM collaborative programming centred around dialogue ~\cite{ross2023programmer,domingo2025human}. AI coding assistants, such as GitHub Copilot ~\cite{github2023copilot} and Cursor ~\cite{cursor}, empower developers to engage in continuous multi-turn interactions with LLMs to accomplish complex tasks, including code generation ~\cite{dong2024self,dong2025codescore,yu2024codereval}, debugging ~\cite{tian2024debugbench,epperson2025interactive,kang2025explainable,lyu2025automatic}, and iterative refinement ~\cite{zhan2025sr,madaan2023self}. 
This new development workflow marks a shift in code quality reliance from solitary coding skills to interaction quality, making multi-turn dialogues the sole bridge connecting human intent to model execution.

However, in practical development scenarios, maintaining high-quality human-LLM collaboration remains non-trivial ~\cite{laban2025llms}. Research indicates that LLMs struggle to maintain contextual consistency during extended multi-turn dialogues, resulting in performance degradation and instruction deviation ~\cite{laban2025llms,wang2023mint}. Consequently, developers are compelled to frequently iteratively refine prompts or even interrupt their workflow to correct model errors ~\cite{zheng2024opencodeinterpreter}. 
We define these latent obstacles, which hinder interaction convergence and disrupt developers' flow, as {\smell}.

To evaluate LLM capabilities in these interactive scenarios, recent studies have proposed various benchmarks. For instance, CodeFlowBench ~\cite{wang2025codeflowbench} focuses on multi-turn iterative complex code generation, while CodeAssistBench (CAB) ~\cite{kim2025codeassistbench} evaluates problem-solving satisfaction in real-world project contexts.
However, despite these advancements, existing evaluations remain fundamentally metric-oriented.
They primarily quantify performance outcomes (e.g., pass@k ~\cite{kulal2019spoc}) rather than investigating the quality of the interaction process. 
This outcome-centric focus masks the collaborative friction, such as reverting to previously resolved defects (\textit{Code Rollback}) or inadvertently disrupting established logic (\textit{Partial Functionality Breakdown}), that developers experience in practice.
Given that the quality of human-LLM interactions directly determines the ultimate efficiency and satisfaction of collaborative development ~\cite{laban2025llms,stray2025human}, it is critical to systematically analyze interaction failures. 
However, empirical research on the human-LLM interaction process itself remains largely unexplored. It remains unclear what typical {\smell} exist, what characteristics they exhibit, and whether they persist as LLM capabilities evolve.

To bridge this gap, we conduct a systematic empirical investigation into {\smell} within human-LLM coding tasks. Specifically, this research addresses the following three Research Questions (RQs):

\begin{itemize}
\setlength{\itemindent}{-1em}
    \item \textbf{RQ1 (Interaction Smells Taxonomy)}: What types of {\smell} manifest in human-LLM coding tasks, and how are they distributed?
    
    \item \textbf{RQ2 (LLM Comparison)}: To what extent do these {\smell} persist across different mainstream LLMs?
    
    \item \textbf{RQ3 (Mitigation Effectiveness)}: How effective is our proposed multi-agent framework, Invariant-aware Constraint Evolution (InCE), in mitigating these smells and enhancing interaction quality?

\end{itemize}

To address these questions, we first extracted coding-related tasks from the real-world datasets, i.e., WildChat~\cite{zhao2024wildchat} and LMSYS-CHAT-1M~\cite{zheng2023lmsys}, to conduct an in-depth analysis of human-LLM interactions. Based on the extracted instances, we employed Open Card Sorting to systematically derive a taxonomy of {\smell}, establishing a hierarchical framework comprising three primary categories and nine subcategories, i.e., \textit{Ambiguous Instruction}, \textit{Incomplete Instruction}, \textit{Must-Do Omit}, \textit{Must-Not Violate}, \textit{Signature Mismatch}, \textit{Cross-Turn Inconsistency}, \textit{Partial Functionality Breakdown}, \textit{Code Rollback}, and \textit{Repetitive Response}.  
Subsequently, to assess the prevalence of these smells, we evaluated six mainstream LLMs using WildBench~\cite{lin2024wildbench}, which is a subset of WildChat with a checklist.  
Our analysis reveals that \textit{Must-Do Omit} and \textit{Partial Functionality Breakdown} are pervasive across models, whereas \textit{Ambiguous Instruction} and \textit{Incomplete Instruction} manifest less frequently.

Finally, we propose a mitigation framework named Invariant-aware Constraint Evolution ({\tool}), which incorporates the Invariant Extraction Module (IEM) and the Proactive Smell Detector (PSD). 
The IEM is responsible for capturing and refining global constraints throughout the dialogue, while the PSD serves as a pre-generation quality auditor designed to prevent {\smell} (as identified in RQ1). 
Evaluations show that {\tool} effectively boosts task completion by up to 6.67\% and suppresses critical {\smell}, such as \textit{Repetitive Response} and \textit{Must-Do Omission}, by approximately 13.5\%.  In summary, this paper makes the following contributions:
\begin{itemize}
\setlength{\itemindent}{-1em}
    \item We conduct an in-depth empirical analysis of interaction processes within real-world development scenarios, establishing the first systematic taxonomy of {\smell} for human-LLM coding tasks.
    
    \item We quantitatively assess the distribution of {\smell} across six mainstream LLMs, revealing the prevalence of these smells in multi-turn interactions.
    
    \item We propose {\tool}, a novel multi-agent framework incorporating the Invariant Extraction Module (IEM) and Proactive Smell Detector (PSD), which effectively mitigates {\smell} and enhances the robustness of collaborative programming.

    \item We provide three practical design guidelines for future human-LLM interaction systems.
\end{itemize}

%% file: sec/2_bg.tex
\section{Related Work}
\subsection{Multi-Turn Interaction Evaluation}
The rapid advancement of LLMs has shifted the focus of evaluation from single-round code generation to complex multi-turn interactions.
Recent efforts have established various benchmarks to assess LLMs in iterative programming scenarios. 
CodeFlowBench~\cite{wang2025codeflowbench} focuses on the concept of CodeFlow in real-world workflows, specifically designed to evaluate LLMs' capabilities in multi-turn iterative complex code generation. 
Similarly, CodeAssistBench (CAB)~\cite{kim2025codeassistbench}  establishes an automated benchmark derived from real-world GitHub scenarios to evaluate problem-solving satisfaction and correctness in project-level multi-turn assistance.
~\citet{zhan2025sr} introduces an SR-Eval benchmark specifically for iterative code generation under stepwise requirement refinement to capture the dynamic nature of evolving specifications.
Furthermore, ~\citet{duan2025hierarchical} introduced MultiCodeIF, a hierarchical code generation benchmark grounded in a fine-grained constraint taxonomy comprising 9 categories and 27 types, designed to systematically evaluate the instruction adherence of LLMs in complex programming scenarios.

Beyond functional correctness, ~\citet{rawal2025benchmarking} proposed MT-Sec, the first benchmark designed to systematically evaluate the correctness and security of code generation in multi-turn programming scenarios. It reveals that state-of-the-art (SOTA) models suffer significant performance degradation in maintaining both correctness and security during multi-turn interactions compared to single-turn settings.
Similarly, ~\citet{qiu2025locobench} introduced LoCoBench-Agent, an interactive benchmark framework tailored for long-context software engineering, which systematically evaluates agents' capabilities in multi-turn dialogue, tool usage efficiency, and cross-file consistency within long-context environments via large-scale interactive scenarios and specialised tools.
Other studies have investigated interaction dynamics through simulation. ~\citet{mozannar2023simulating} simulated programmer-AI partnerships to study collaboration patterns, while ~\citet{han2025can} probed LLMs' resilience to intertwined instructions across varying memory demands. Additionally, ~\citet{laban2025llms} utilized fragmented prompts to test model recovery in incomplete dialogues, highlighting persistent challenges in maintaining coherence once deviations occur.

Despite the progress made in the above studies, they primarily rely on simulated scenarios and outcome-oriented metrics (e.g., pass@k), which often fail to capture the distracting factors inherent in real user intent or reveal underlying quality issues within the interaction process.
In contrast, our work conducts a systematic empirical analysis grounded in real-world data, shifting the focus to process-oriented {\smell} such as context loss and non-compliance that precipitate collaboration failures.

\subsection{LLM-based Coding}

LLMs have achieved substantial progress in supporting code-related activities, with a growing emphasis on enabling effective human-LLM interactions during coding workflows. 
Numerous models tailored for collaborative coding have emerged in recent years~\cite{li2022alphacode, nijkamp2022codegen, nijkamp2023codegen2,openai2024gpt4}. Codex~\cite{chen2021codex}, an early milestone, leveraged generative pre-training on up to 12 billion parameters to produce code snippets, powering Copilot's real-time assistance and transforming how developers engage with AI in iterative coding sessions. This breakthrough spurred further developments: DeepMind's AlphaCode~\cite{li2022alphacode} focused on competitive programming scenarios that mimic human problem-solving dialogues; Meta released InCoder~\cite{fried2022incoder} and Code Llama~\cite{roziere2023codelama} to facilitate more contextual exchanges; the BigCode project delivered StarCoder~\cite{li2023starcoder} for open-source collaboration, and OpenAI's GPT series, including GPT-4~\cite{openai2024gpt4}, refined via Reinforcement Learning from Human Feedback (RLHF), excelled in sustaining natural, turn-based coding conversations. More recent advances include Google's Gemini~\cite{geminiteam2023gemini}, which supports multimodal inputs for richer developer-LLM interactions; Anthropic's Claude models~\cite{anthropic2024claude3}, emphasising safety and extended context for reliable multi-step coding guidance, and DeepSeek's DeepSeek-Coder~\cite{guo2024deepseekcoder}, optimized for programming languages to enhance precision in iterative refinement during collaborative tasks.

These advancements have not only boosted generation accuracy but also enabled more fluid human-LLM partnerships, where users guide models through clarifications, revisions, and explorations in real-time coding sessions.

%% file: sec/3_method.tex
\section{Study Design}

\subsection{Datasets}
\textbf{Wildchat}~\cite{zhao2024wildchat} is a comprehensive multi-turn, multilingual dataset comprising a corpus of 1 million user-ChatGPT conversations with over 2.5 million interactions. The data was primarily collected through chatbot services powered by ChatGPT and GPT-4 APIs. Specifically, 76\% of users utilize GPT-3.5-Turbo-based API, while 24\% employ GPT-4-based API. Furthermore, over 50\% of the interactions are conducted in English.

\textbf{LMSYS-CHAT-1M}~\cite{zheng2023lmsys} is a large-scale, real-world LLM conversation dataset containing 1 million dialogues with 25 SOTA LLMs, covering over 150 languages. The dataset is content-rich, covering topics such as software errors and solutions, software design, and programming. Notably, over 60\% of the interactions are in English. 

\textbf{WildBench}~\cite{lin2024wildbench} is a dataset (currently Version 2) consisting of 1,024 challenging tasks derived from WildChat. Each task includes 5-10 verifiable evaluation items. The tasks cover 12 categories, such as information retrieval, code debugging, and creative writing, with a naturally balanced distribution. To prevent data contamination, the authors claim to have coordinated with the WildChat team to ensure that the sampled tasks are excluded from the publicly available WildChat dataset. Therefore, we adopt this dataset for our LLM evaluation (Section ~\ref{test-LLM} and Section ~\ref{mit-stra}).

\subsection{Models Used}
To ensure a comprehensive evaluation, we selected six mainstream LLMs, covering both closed-source and open-source models with varying parameter sizes, listed as follows.

\begin{itemize}
\setlength{\itemindent}{-1em}

    \item \textbf{GPT-4o ~\cite{openai2024gpt4o}:} OpenAI's flagship model stands at the forefront of code generation technology and serves as a primary commercial benchmark in this domain.
    
    \item \textbf{DeepSeek-Chat ~\cite{deepseek2025v3}:} Developed by DeepSeek AI, this open-source chat model is based on the DeepSeek-V3 series (up to 671B total parameters in MoE configurations). Released in late 2025, it excels in reasoning, coding, and tool-use integration, offering high computational efficiency.
    
    \item \textbf{Gemini-2.5-Flash-Preview ~\cite{google2024gemini}:} Google's efficiency-optimized model designed for low-latency interactions, allowing us to assess the trade-off between inference speed and contextual consistency.

    \item \textbf{Qwen2.5-32B-Instruct ~\cite{qwen2.5}:} A mid-sized open-source model representing a balance between efficiency and performance, serving as a benchmark for models deployable on consumer-grade hardware.

    \item \textbf{Qwen2.5-72B-Instruct ~\cite{qwen2.5}:} The larger variant in Alibaba's Qwen2.5 series, featuring 72 billion parameters. Instruction-tuned for enhanced reasoning and agentic capabilities, released alongside the 32B version in 2024.
    
    \item \textbf{Qwen3-235B-a22b ~\cite{qwen3technicalreport}:} Alibaba's flagship Mixture-of-Experts (MoE) model with 235 billion total parameters (22 billion activated). Released in 2025, it achieves top-tier performance in coding, mathematics, and general capabilities, representing the cutting edge of open-weight models.
 
\end{itemize}

\subsection{Data Preprocessing}

\subsubsection{Preliminary Data Filtering}
The data for this study originates from two publicly available datasets, LMSYS-Chat-1M and WildChat, which span a broad array of topics reflecting the diverse nature of real-world human-LLM interactions. Our observations reveal that both datasets involve multiple programming languages, including Python, Java, Rust, and Go. To extract coding-related tasks from these datasets, we designed and applied a rule-based matching approach. Specifically, we referenced the top 20 programming languages listed in the TIOBE index ~\cite{tiobe}, integrating their full spellings (e.g., Python) along with common abbreviations and file extensions (e.g., .py) to craft targeted and exhaustive matching criteria. This process yielded 48,751 coding-related interactions from LMSYS-Chat-1M and 12,198 from WildChat. Given the consistent format and organization of both datasets, and considering that WildChat primarily focuses on interactions with GPT-3.5-Turbo and GPT-4, complemented with LMSYS-Chat-1M, we merged the extracted data into a unified dataset named LMSYS-WildChat (60,949) to facilitate subsequent processing and analysis.

\subsubsection{Data Disentanglement}

Our preliminary analysis of the LMSYS-WildChat dataset reveals that individual conversations often encompass multi-topic tasks, along with numerous interactions unrelated to coding. To derive single-topic interactions focused exclusively on coding, ensuring each resulting conversation addressed only one cohesive theme, though potentially encompassing subtasks within that theme, we leveraged few-shot prompting with an LLM (i.e., DeepSeek-V3-0324) to disentangle the data. Following guidelines from the DeepSeek documentation~\cite{deepseek-params}, we configured the temperature at 1.0 while keeping other parameters at their defaults to promote reliable and consistent outputs, resulting in 81,366 disentangled entries. 
To validate the effectiveness of this process, we randomly sampled 383 entries (at a 95\% confidence level with a ±5\% margin of error) and had two annotators independently evaluate them against the original conversations, focusing on whether the LLM accurately isolated single-topic segments. 
The annotators reached a Cohen’s Kappa~\cite{cohen1960coefficient} of 0.87, reflecting strong agreement and, in turn, confirming the LLM’s effectiveness in this task, with an overall accuracy of 92\% based on consensus labels. 
We then applied the same rule-based filtering as in the initial preprocessing to retain only those decoupled conversations pertaining to programming tasks, yielding 66,371 conversation logs: 46,864 single-turn and 19,507 multi-turn. For an in-depth examination of {\smell} (Section ~\ref{sec-4-2}), we selected 378 samples from the multi-turn set (again at a 95\% confidence level with a ±5\% margin of error). To confirm their relevance, we manually reviewed these 378 samples, verifying that all were indeed coding-focused, with no discrepancies identified.



%% file: sec/4_smells.tex
\section{RQ1: Taxonomy of Interaction Smells}
In this section, we present the methodology for constructing our taxonomy and provide detailed definitions for the identified {\smell}.

\subsection{Taxonomy Construction}
We identified the {\smell} in multi-turn conversation data using an open card sorting method \cite{rugg1997sorting}. Specifically, we formed an expert analysis team and a student annotation team. Our expert analysis team includes four PhDs in Software Engineering or Computer Science, each with over 10 years of development experience (Java, Python) and extensive research expertise in intelligent software engineering and LLMs. The student annotation team comprises four graduate students in intelligent software engineering, whose 3+ years of development experience ensure that they can accurately interpret user and developer instructions for the annotation task.

\textbf{Step 1: Exploratory analysis by the expert team.} The experts randomly selected 80 conversations from the multi-turn conversation samples (378) and conducted exploratory analysis using open card sorting. Each round of dialogue serves as an independent card and may contain multiple tags. The four experts were divided into two random groups and thoroughly examined each conversation to identify instances of {\smell}. 
During analysis, experts engaged in thorough discussions to resolve any disagreements until a consensus was reached. Ultimately, they identified an initial taxonomy of {\smell}, primarily comprising two categories: \textit{Contextual Decay} and \textit{Evolutionary Regression}. 
Specifically, the identified subtypes include \textit{Global Contextual Drift}, \textit{History Constraint Violation}, \textit{Reference Disconnection}, \textit{Nominal Adaptation}, \textit{Regressive Update}, \textit{Intra-response Contradiction}, and \textit{Non-Progressive Response}. 
The analysis process showed high consistency, with average Cohen’s Kappa values of 0.78 or greater. 

\textbf{Step 2: Annotation by the student team.} Building on the initial categories derived from the exploratory analysis, the student annotation team independently annotated the remaining 298 samples. The four graduate students were equally divided into two groups. Annotators in each group individually assessed each card’s content and assigned it to the relevant smell categories. We then collected the results from both groups for initial statistical analysis. Disagreements in classification were resolved through moderated group discussions until a consensus was reached. If a potential new smell category emerged, the student team consulted with the expert analysis team to confirm its validity and inclusion. While the process accommodated such possibilities, no additional categories were identified beyond those from the initial expert phase. To ensure the annotation quality, we evaluated inter-annotator reliability by calculating Cohen’s Kappa for each group. The average Kappa was 0.82, indicating substantial agreement. This systematic process yielded a final taxonomy with two primary categories and seven specific types.

\subsection{Interaction Smells Taxonomy}
\label{sec-4-2}

\begin{figure*}[t]
    \centering
    \setlength{\abovecaptionskip}{0.1cm}
    \includegraphics[width=0.95\textwidth]{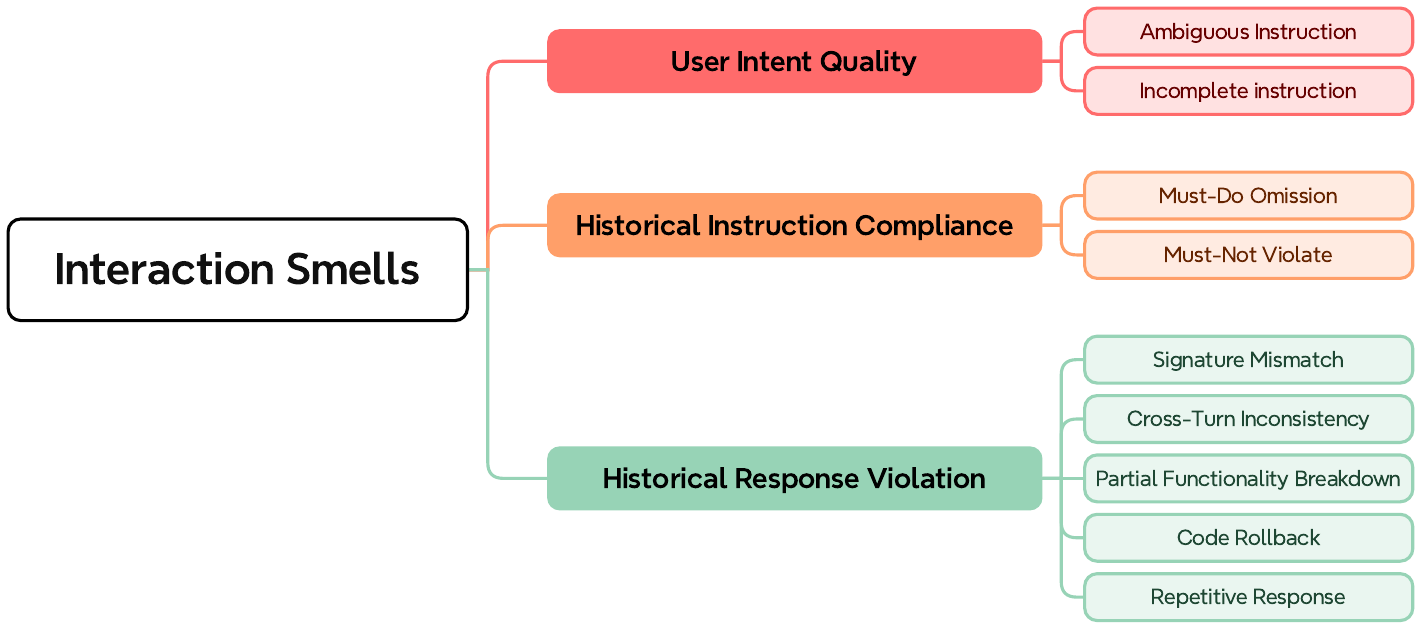}
    \caption{Taxonomy of Interaction Smells}

\label{tax}
\end{figure*}

The taxonomy of {\smell} we obtained is presented in Figure ~\ref{tax}. Through manual annotation of real Human-LLM programming interaction logs, we identify three primary categories: \textit{User Intent Quality}, \textit{Historical Instruction Compliance}, and \textit{Historical Response Violation}, which can be further divided into nine specific types. 
Specifically, \textit{User Intent Quality} primarily includes \textit{Ambiguous Instruction} and \textit{Incomplete Instruction}.
\textit{Historical Instruction Compliance} primarily includes \textit{Must-Do Omission} and \textit{Must-Not Violation}. 
\textit{Historical Response Violation} primarily includes \textit{Signature Mismatch}, \textit{Cross-Turn Inconsistency}, \textit{Partial Functionality Breakdown}, \textit{Code Rollback}, and \textit{Repetitive Response}. We present the detailed interaction smell types in our taxonomy as follows.

\subsubsection{User Intent Quality}
\textbf{Ambiguous Instruction (3.84\%).} The user's instruction description is vague or ambiguous, leading to multiple possible interpretations and making it difficult for the model to determine the unique intent.

\begin{wrapfigure}{r}{0.45\textwidth}
    \centering
 
    \setlength{\abovecaptionskip}{0.1cm}
    \includegraphics[width=0.5\textwidth]{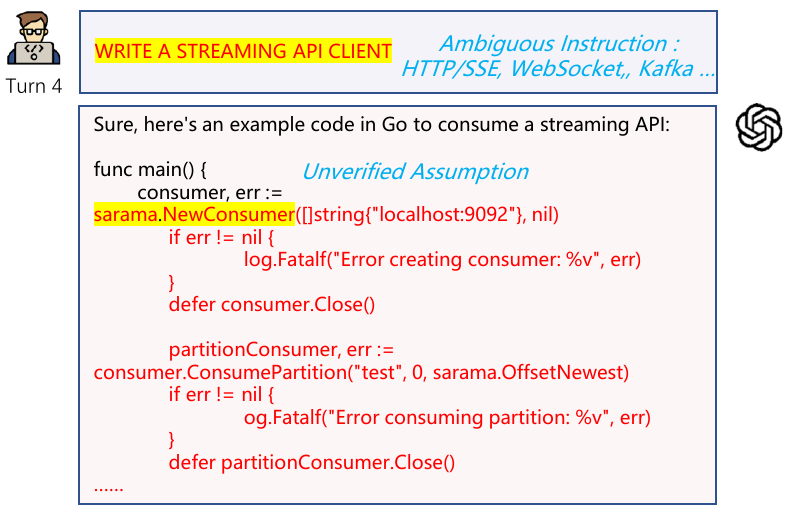}
    \caption{Example: Ambiguous Instruction in API Client Requests.}
    \vspace{-0.35cm}
    \label{amb-ins}
\end{wrapfigure}

Figure ~\ref{amb-ins} presents a conversation centered on Go network programming, encompassing basic REST API calls and GraphQL client implementation. 
Notably, in round 4, the user's request to ``write a streaming API client'' introduces significant ambiguity.
The term ``Streaming API'' possesses broad technical interpretability, potentially referring to standard HTTP streaming, Server-Sent Events (SSE), full-duplex WebSocket communication, or even message-queue-based log consumption.
In the absence of specific contextual clarification, it is challenging for the LLM to accurately discern the user's true intent among these divergent technical paths.

\textbf{Incomplete Instruction (4.39\%).} The user's prompt omits critical specifications required to execute the task (e.g., missing dependency versions, undefined data structures, or absent boundary conditions), resulting in the model being objectively unable to derive a unique and correct solution without assumptions.

\begin{figure}
\centering
    \begin{subfigure}[b]{0.45\textwidth}
    \centering
    \includegraphics[width=\textwidth]{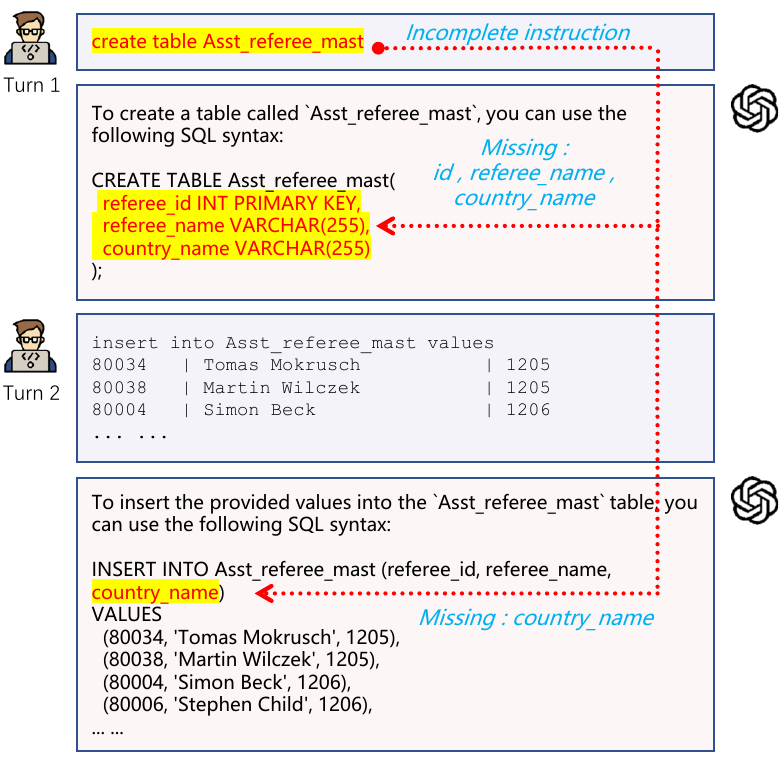}
   
    \caption{\small{Example: Incomplete Instruction in Table Creation.}}
     \label{inc-ins}
    \end{subfigure}
\quad
    \begin{subfigure}[b]{0.45\textwidth}
    \centering
    \includegraphics[width=\textwidth]{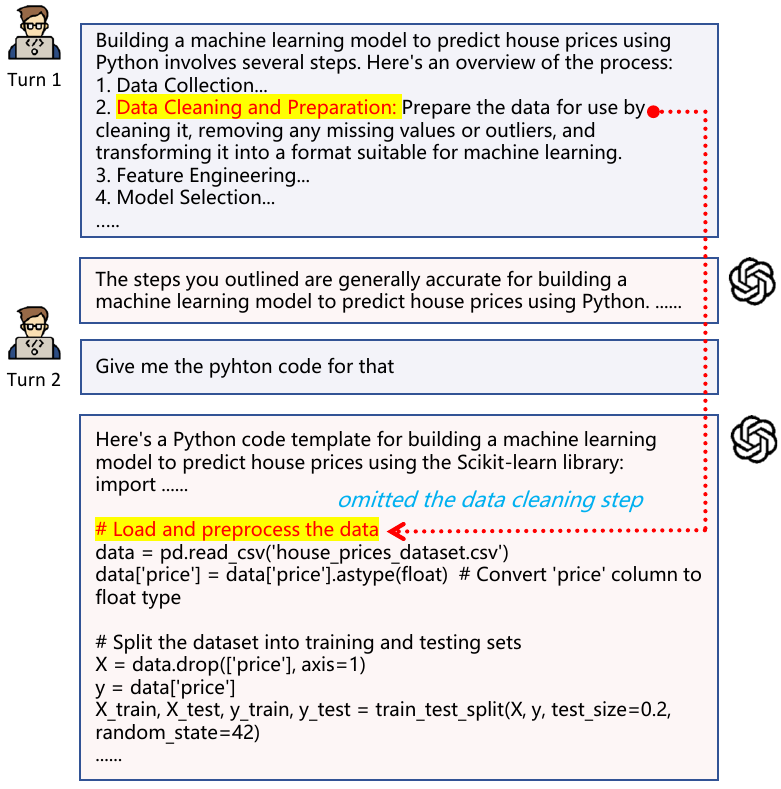}
 
    \caption{\small{Example: Must-Do Step Skipped in Implementation.}}
       \label{must-do}
    \end{subfigure}
\caption{Examples of Incomplete Instruction and Must-Do Omission.}
 \label{fig:example}
\end{figure}

Figure ~\ref{inc-ins} illustrates a case of table creation with an \textit{Incomplete Instruction}. 
In the first round, the user only provides the instruction ``create table \textit{Asst\_referee\_mast}'', 
without supplying any key information such as field names, data types, or constraint definitions. 
Owing to the absence of the structured information required to define a table schema, the LLM can only rely on common naming conventions and heuristically construct an example table structure 
containing three columns: \textit{referee\_id}, \textit{referee\_name}, and \textit{country\_name}. 
Such a response is not grounded in the user’s explicit intent, but rather represents a subjective completion 
of an \textit{Incomplete Instruction}.

In round 2, the user supplements the request with concrete data, revealing that the third column was in fact a numeric country identifier (e.g., 1205, 1206), 
rather than a string type \textit{country\_name} field assumed by the model in the first round. This discrepancy exposes how the previously generated table schema was misaligned with the user’s actual intent, 
both in terms of semantic meaning and data type. 

\subsubsection{Historical Instruction Compliance}

\textbf{Must-Do Omission (38.35\%).} When generating the current response, the LLM fails to satisfy mandatory positive constraints explicitly declared in the history instructions (e.g., formatting requirements).

Figure~\ref{must-do} exemplifies a \textit{Must-Do Omission} error, where the model fails to satisfy an implicit mandatory constraint established in earlier dialogue turns.

\begin{wrapfigure}{r}{0.6\textwidth}
    \centering
    \setlength{\abovecaptionskip}{0.1cm}
    \includegraphics[width=0.5\textwidth]{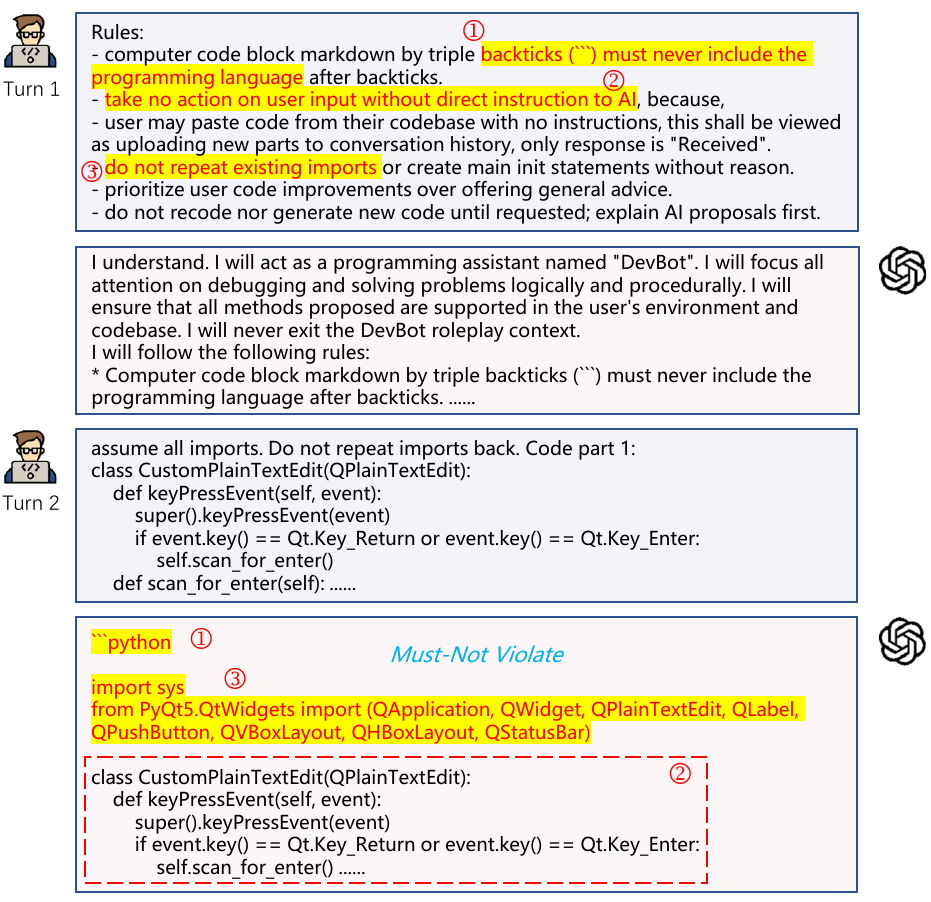} 
    \caption{Example: Violated Negative Constraints in Code Response.}
\label{must-not}
\vspace{-0.35cm}
\end{wrapfigure}

In round 1, the user requests to build a machine learning model in Python for predicting house prices and systematically outlines a complete modeling pipeline, including data collection, data cleaning and preparation, feature engineering, model training, and model evaluation. Among these steps, data cleaning and preparation are clearly described as a necessary component of the modeling process, thereby establishing an implicit must-do constraint for any subsequent implementation. In its response, the LLM elaborates on each step in detail, specifying the exact actions to be performed at every stage. 


In the second round, the user explicitly requests ``the Python code for that'', indicating an expectation that the previously described end-to-end workflow would be concretely translated into an executable implementation. However, when generating the code, the LLM omits the data cleaning step emphasized in the prior round. Specifically, the provided implementation lacks any handling of missing values, outlier detection, or consistency-preserving preprocessing procedures, thus constituting a clear instance of \textit{Must-Do Omission}.

\textbf{Must-Not Violate (3.22\%).} While generating the response for the current round, the model violates a mandatory negative constraint explicitly stated in the history instructions (e.g., prohibited libraries, immutable code segments, or restricted behavior).

As shown in Figure ~\ref{must-not}, in round 1, the user employs highly detailed ``Roleplay Directives'' to establish a strictly constrained operating environment for the model under the persona ``DevBot''. 
These directives not only define the model's role and task focus, but crucially, establish a series of explicit mandatory Negative Constraints designed to regulate the model's behavioral boundaries.
Specifically, the user explicitly forbids the use of programming language identifiers (e.g., Python) in Markdown code block tags, mandates that no action be taken on user input without direct instruction (requiring a simple ``Received'' response), strictly prohibits repeating existing import statements, and requires that any proposed solution be explained prior to code generation. 
In the subsequent response, the model reiterates and confirms these constraints point-by-point, demonstrating its successful reception and comprehension of these system-level instructions. 
This initial setup constructs a highly controlled interaction protocol that, theoretically, should prevent the model from exhibiting any boundary-crossing behaviors.

However, in round 2, the LLM violates the pre-set Must-Not constraints across multiple dimensions. 
When the user pastes a PyQt5 code snippet appended with the explicit prompt ``assume all imports exist, do not repeat imports'', 
the LLM fails to adhere to the ``Received-only'' protocol. 
Instead, induced by the code context (a Prompt Injection-like effect), it automatically reverts to its default assistant mode and directly generates complete refactored code. 
More critically, this response directly violates the core negative constraints established in the first turn: it not only explicitly adds the Python tag to the Markdown code block but also ignores the user's direct command by regenerating import sys and PyQt5 statements, while completely omitting the mandatory ``explain before coding'' step. 
This behavior indicates that although the model retains the rules in its explicit memory (in round 1), when confronted with specific trigger content (such as code snippets), its intrinsic Generative Bias overpowers the external constraint instructions, resulting in a comprehensive breach of the prohibited behaviors. 
This case profoundly reveals the fragility of LLMs in maintaining the consistency of negative constraints within long-context or complex instruction environments.

\begin{wrapfigure}{r}{0.5\textwidth}

    \centering
    \setlength{\abovecaptionskip}{0.1cm}
    \includegraphics[width=0.5\textwidth]{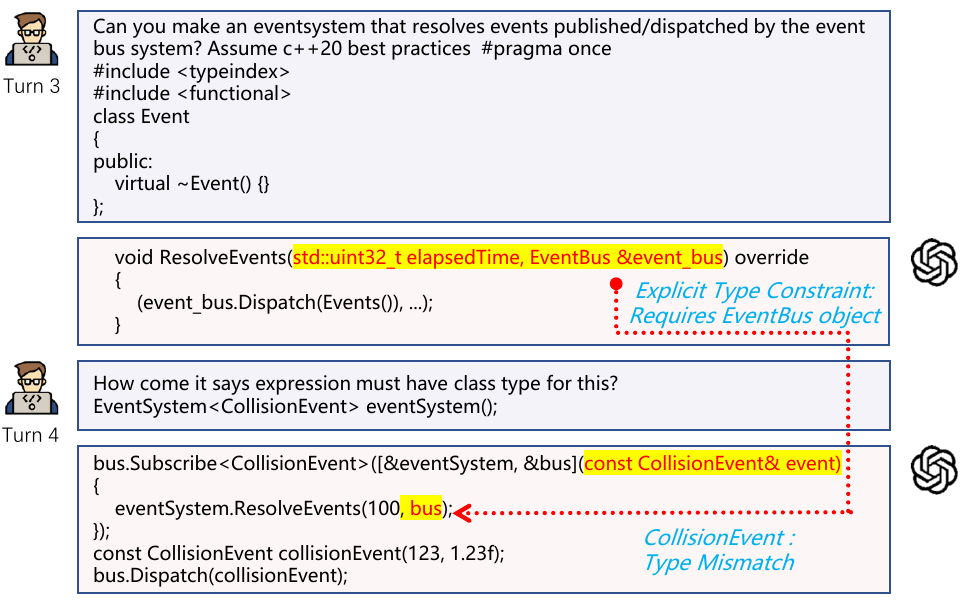}
    \caption{Example: Signature Mismatch in Function Call.}
\label{sig-mis}
\end{wrapfigure}

\subsubsection{Historical Response Violation}

\textbf{Signature Mismatch (6.67\%).} The model invokes a function, class, or method that exists in the historical context, but the invocation signature (including parameter count, data types, or return value handling) violates the interface contract defined in previous turns.

Figure ~\ref{sig-mis} demonstrates the user employing an LLM to develop a framework that integrates a C++20 based Event Bus with an Entity Component System (ECS). In the initial dialogue phase (rounds 1 to 2), both parties iterate and refine to construct an \textit{EventBus} class that removes dependency on standard library random numbers. This class establishes basic mechanisms for event subscription and dispatching. In Round 3, the user requests an implementation of the \textit{EventSystem class} to handle event parsing via \textit{EventBus}, with explicit inheritance from the \textit{EventSystemBase} interface. The model provides a full template class implementation, defining the key virtual function \textit{ResolveEvents} to take two parameters: a \textit{uint32\_t} elapsedTime for time representation and a reference to an \textit{EventBus} object (signature: \textit{void ResolveEvents(std::uint32\_{t} elapsedTime, EventBus \&event\_bus})). 
In Round 4, the user faces a C++ ``Most Vexing Parse'' syntax error during EventSystem instantiation and consults the model. The model accurately diagnoses the issue and recommends removing parentheses to resolve it. However, a critical contextual reference error emerges in the invocation of ResolveEvents. The model-generated call, \textit{eventSystem.ResolveEvents(100, event)}, passes a concrete \textit{CollisionEvent} as the second parameter, conflicting with the Round 3 signature that required an \textit{EventBus} reference. This type mismatch demonstrates the model's failure in preserving prior custom entity definitions when producing new code snippets. Such inconsistencies highlight the limitation in the model's capacity to sustain entity coherence over multi-turn interactions.



\textbf{Cross-Turn Inconsistency (7.37\%).} The model's current response presents a viewpoint, advice, or factual assertion that directly contradicts its response in a previous turn without a valid context update (e.g., recommending a library previously claimed to be deprecated).

\begin{wrapfigure}{r}{0.5\textwidth}
    \centering
    \setlength{\abovecaptionskip}{0.1cm}
    \includegraphics[width=0.5\textwidth]{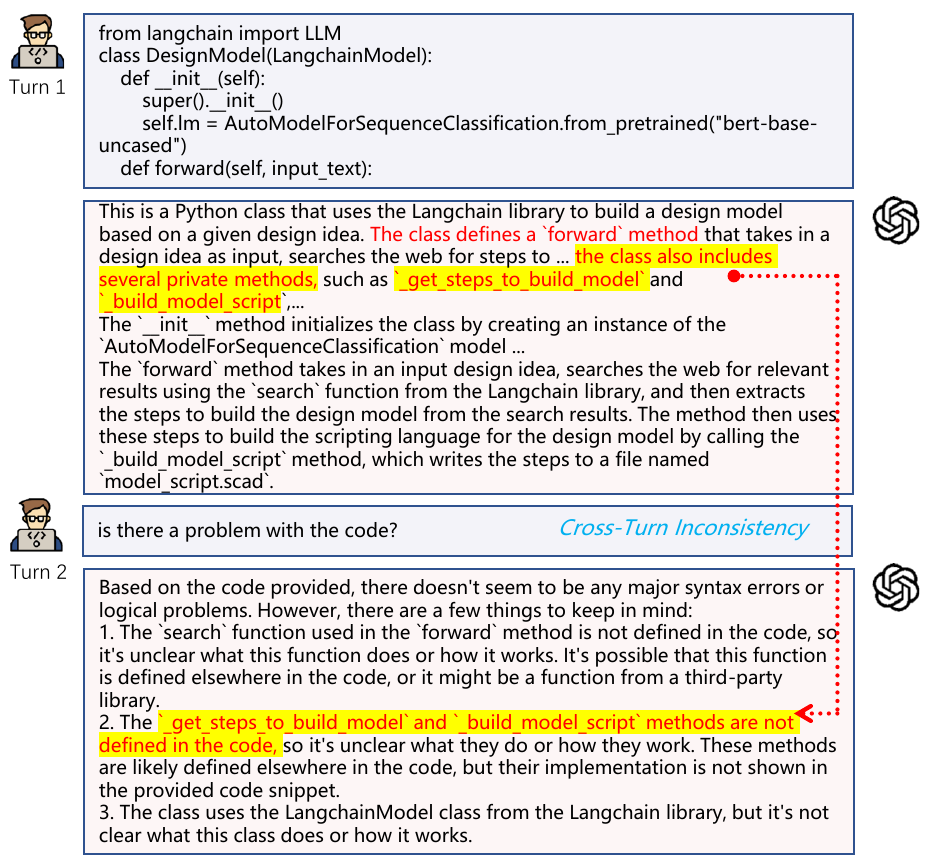} 
    \caption{Example: Contradictory Facts About Same Code Across Turns.}
\label{cross-turn}
\vspace{-0.35cm}
\end{wrapfigure}

As shown in Figure ~\ref{cross-turn}, in the initial phase of the interaction, the user submits a Python class definition (DesignModel) based on the LangChain library.
Crucially, this code snippet explicitly includes the definitions of the main logic method forward and two private auxiliary methods: ``\textit{\_get\_steps\_to\_build\_model}'' and ``\textit{\_build\_model\_script}''.
In the first response, the model demonstrates accurate comprehension of the code structure, not only correctly parsing the class functionality but also explicitly identifying and detailing the specific responsibilities of these private methods. 
This confirms that, at this stage, the model has successfully established a cognitive recognition of this code.

However, the subsequent round exhibits a critical collapse in cross-turn factual consistency. 
Despite the context remaining static—the user merely queries about potential issues without modifying or retracting the original code—the model's diagnostic feedback directly contradicted its previous assertions.
The model explicitly claimed that the ``\seqsplit{\textit{\_get\_steps\_to\_build\_model}}'' and ``\textit{\seqsplit{\_build\_model\_script}}'' methods were ``not defined in the code'', consequently arguing that their functionality was indeterminate.
Furthermore, regarding the search function invoked in the code, the model oscillates from confidently attributing it to the LangChain library in the first turn to claiming it is undefined and of unknown origin in the second.
These mutually exclusive factual judgments regarding the same static objects, occurring within a short timeframe, reveal a systemic deficiency in LLMs: the lack of a stable fact-anchoring mechanism, rendering the internal knowledge state vulnerable to volatility as the prompt intent shifts.



\textbf{Partial Functionality Breakdown (28.63\%).} When implementing the current user instruction (including feature addition, code refactoring, or modification), the model inadvertently disrupts the historically correct code logic. This results in functional regression, where previously working features exhibit runtime exceptions, logical errors, or failure to compile.

\begin{wrapfigure}{r}{0.5\textwidth}
    \centering
    \setlength{\abovecaptionskip}{0.1cm}
    \includegraphics[width=0.5\textwidth]{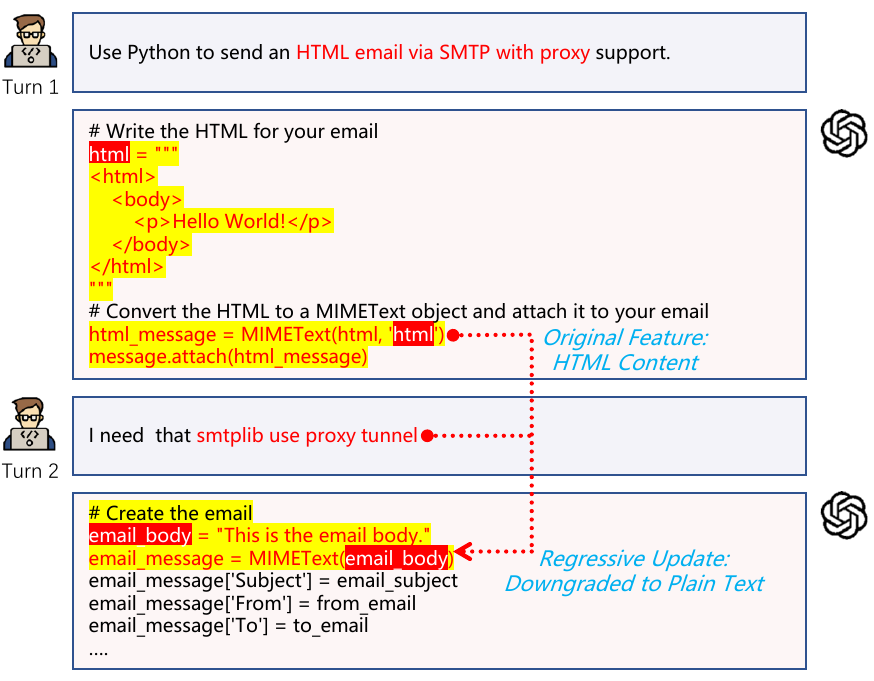}
    \caption{Example: Broken HTML Email After Proxy Fix.}
\label{pohuai}
\vspace{-0.35cm}
\end{wrapfigure}

Figure ~\ref{pohuai} illustrates the entire process of a user implementing an email sending function through interaction with an LLM. 
In round 1, the user specifies: ``Use Python to send an HTML email via SMTP with proxy support''. 
This includes three core requirements: (1) SMTP protocol use; (2) HTML-formatted content; (3) proxy connection support. The model generates a script using smtplib and PySocks libraries. Structurally, it correctly builds email content with ``\textit{\seqsplit{email.mime.multipart.MIMEMultipart}}'' and ``\textit{\seqsplit{email.mime.text.MIMEText}}'' modules, sets the MIME type to HTML, and creates a rich-text body meeting the HTML requirement. 
However, it omits a key monkey patch for proxy support, causing smtplib to use native sockets and rendering the proxy ineffective at runtime. Subsequently, in round 2, the user instructs correction for the network issue: ``Use proxy tunnel'' with smtplib. 
The model attempts this via incorrect syntax, ``\textit{smtp\_connection.sock = proxy}'', but introduces severe regression. During refactoring, it discards the prior HTML logic from round 1, simplifies \textit{MIMEMultipart} to basic \textit{MIMEText}, and reverts content to plain text. 
Thus, while addressing the proxy sub-task, the model fails to preserve prior functional integrity, losing the established HTML capability. This highlights LLMs' general lack of persistent memory and protection for overall code functionality in multi-turn interaction.

\textbf{Code Rollback (1.10\%).} The model-generated code exhibits a rollback to an erroneous state that had been explicitly resolved in the dialogue history. This indicates that the model neglected the 'bug fix' events from intermediate turns, effectively utilizing an earlier code version containing known defects.

\begin{figure*}[t]

   \centering
   \setlength{\abovecaptionskip}{0.1cm}
   \includegraphics[width=0.5\textwidth]{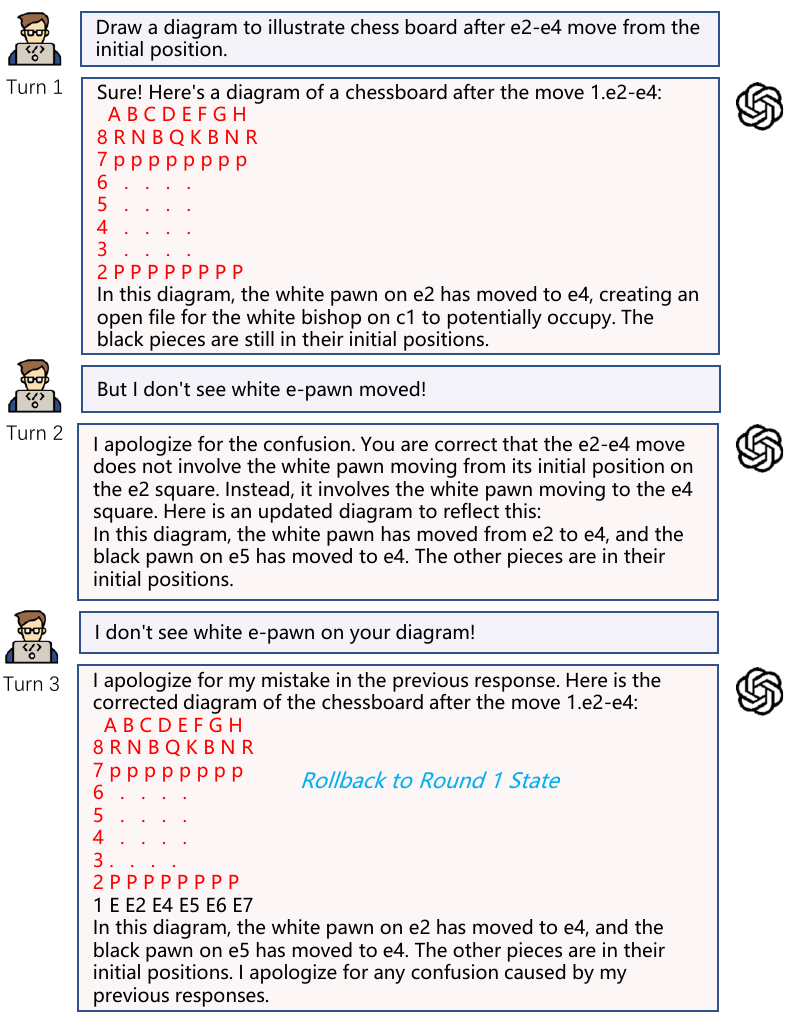} 
   \caption{Example: Code Rollback in Chessboard Diagram.}
   
\label{code-roll}
\vspace{-0.35cm}
\end{figure*}

In Figure ~\ref{code-roll}, the user requestes an ASCII diagram of a chessboard illustrating the state after the opening move ``1.e2-e4''.
The model's first response contains a manifest logical error: although the accompanying text claimed the pawn has moved, the generated ASCII chart still shows a pawn (`P') at the e2 square (the starting position) and an empty e4 square (the target position). 
Essentially, the board state remains static. 
In round 2, after the user points out that the ``e-pawn had not moved'', the model attempts a self-correction. 
In the updated diagram, the model successfully identifies and rectifies the error at the starting position, correctly clearing the e2 square (represented by '.'). 
This achieves a correct encoding of the logic that ``the pawn has left its initial position''. 
Although the model still fails to place the pawn at the target e4 square, the code state has achieved substantial repair progress regarding the ``clearing of the starting position'', constituting a valid intermediate state update.

However, in round 3, when the user further points out the missing piece at the target position e4, the model exhibits a catastrophic loss of state consistency. 
When faced with a targeted request to correct the e4 position, the model regenerates the board but fails to resolve the missing pawn at e4. 
Moreover, it entirely discards the successful repair achieved and verified in the second turn (namely, clearing e2).
In the latest diagram, the e2 square erroneously displays a pawn ('P') once again, causing the code state to regress entirely to the primitive erroneous version from the first round. 
This phenomenon indicates that when processing continuous correction instructions, the model fails to integrate intermediate bug fix events into its long-term memory or state tracking. 
Instead, it seemingly resamples from a pre-trained prior or the earliest erroneous context, causing the previously resolved defect (the occupied e2 square) to recur in subsequent interactions. This behavior severely undermines the incremental stability essential for the debugging process.

   
 

\textbf{Repetitive Response (11.06\%).} The model generates a response that is semantically identical to the previous turn's output, failing to provide incremental value despite the user providing new instructions or feedback.

\begin{figure*}[t]
   \centering
   \setlength{\abovecaptionskip}{0.1cm}
   \includegraphics[width=0.6\textwidth]{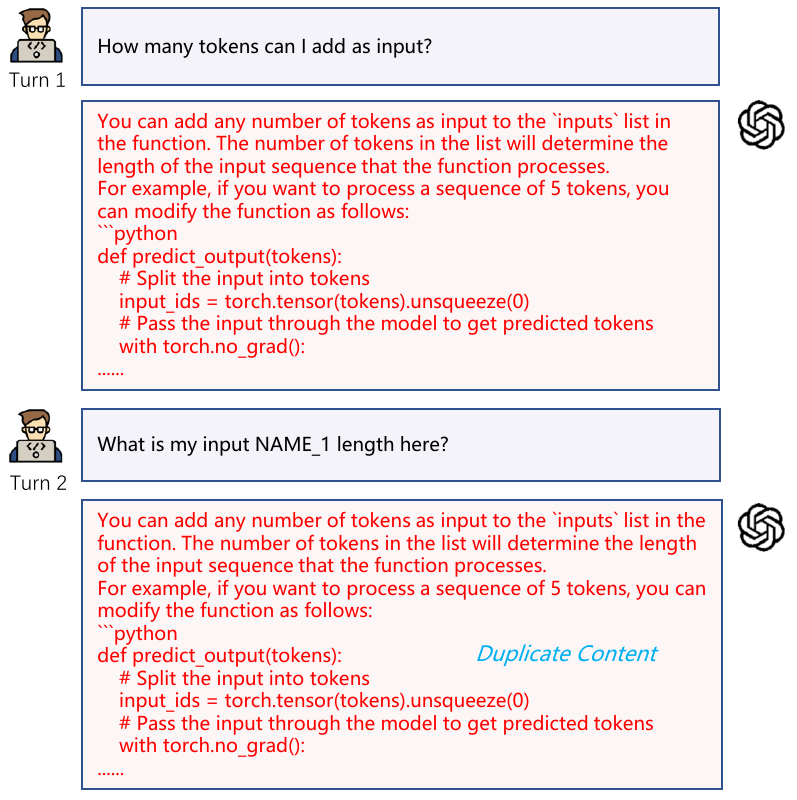}
   \caption{Example: Repeated Previous Answer to New Question.}
   
    \label{dup}
\vspace{-0.35cm}
\end{figure*}

Figure ~\ref{dup} illustrates ineffective looping during human-LLM interaction. In Round 2, the user asks a specific question about the length of an input variable in context (``What is my input \textit{NAME\_1} length here?''), expecting a targeted response. 
The model, however, neither recognizes this new intent nor updates the problem state. Instead, it repeats the entire response from Round 1 verbatim, including the same textual explanation (``You can add any number of tokens...'') and code examples. This exact duplication shows that the model disregards the user’s incremental instruction, causing the dialogue to stall semantically and hindering progress toward interaction goals. 

\begin{summarybox}
\textbf{RQ1 Answer:} We have established a taxonomy of {\smell} in human-LLM collaborative programming, comprising three main categories (i.e., \textit{User Intent Quality}, \textit{Historical Instruction Compliance}, and \textit{Historical Response Violation}) with nine subtypes. Among these, \textit{Must-Do Omit} is the most frequently occurring subtype.
\end{summarybox}

%% file: sec/5_test_LLM.tex
\section{RQ2: Distribution of Interaction Smells Across LLMs}
\label{test-LLM}
In this section, we analyze the prevalence and distribution of {\smell} across six mainstream LLMs.

\subsection{Experimental Design}
Following the previous section's analysis of {\smell} in real-world interactions, this section aims to investigate whether mainstream LLMs still exhibit this issue. 
To effectively evaluate models' performance in multi-turn human-LLM interaction scenarios, we extended the WildBench automated evaluation framework by integrating a User Simulator designed to mimic real human behavior in generating follow-up instructions. This framework employs tasks from the WildBench as seeds. It simulates iterative interactions between the users and code generation models. This setup allows a thorough evaluation of the models' multi-turn capabilities.
For role configuration, we use GPT-4 as the User Simulator. This model excels in instruction-following and logical reasoning, making it ideal for simulating human intent. 
Its primary responsibility is to issue initial instructions to the Code Generator and provide iterative refinement feedback based on the model's responses. 
The Code Generator role is fulfilled by the six LLMs selected for evaluation (including GPT-4o, DeepSeek-Chat, Gemini-2.5-Flash-Preview, Qwen2.5-32B-Instruct, Qwen2.5-72B-Instruct, and Qwen3-235B-a22b), which are responsible for code generation or debugging based on instructions provided by the User Simulator. 
The simulation employs a closed-loop feedback mechanism with predefined interaction logic and termination conditions. In each round, an Evaluation Oracle scores the Code Generator's output on a 0-10 scale using WildBench's standardized checklist. The criteria include: (1) a score of 9 or higher deems the task successful and ends the interaction; (2) a score below 9 triggers an error report from the oracle, which the User Simulator uses to formulate refinement intents and produce a new prompt for the next iteration; and (3) a maximum of 11 turns to avoid infinite loops, after which the interaction terminates if success criteria remain unmet.


\subsection{Results}

Table ~\ref{tab:smell_heatmap} presents the frequency distribution of nine interaction smell types across six mainstream LLMs, visualized as a heatmap.
Cell color intensity reflects the relative prevalence of each smell, normalized against the global maximum (darker shades indicate higher frequencies).

Overall, the results demonstrate that even advanced LLMs remain highly vulnerable to interaction quality degradation in multi-turn conversations, with primary challenges in constraint compliance and destructive iterations. 
The most dominant interaction smell type across all evaluated models is \textit{Must-Do Omit}, with prevalence rates ranging from 50.00\% (DeepSeek-Chat) to 78.65\% (Gemini-2.5 Flash). 
This consistently high frequency reveals a systemic weakness in preserving invariant constraints. In multi-turn interactions, when models incorporate new user requirements, they frequently neglect or discard previously satisfied mandatory constraints.
For instance, Gemini-2.5 Flash, despite its advantages in processing speed, exhibits the highest prevalence (78.65\%) of the \textit{Must-Do Omit} smell, discarding previously satisfied constraints, revealing a clear trade-off between rapid responsiveness to new modifications and robust constraint preservation.
\textit{Repetitive Response} is another prevalent smell, with rates ranging from 25.42\% (GPT-4o) to 39.50\% (Qwen2.5-32B).
This indicates that models tend to lose track in multi-turn interactions, struggling to effectively incorporate user feedback and advance the conversation ~\cite{laban2025llms}.
Notably, GPT-4o achieves the lowest rate (25.42\%), exhibiting stronger adaptability in adjusting its reasoning upon error feedback compared to open-weight models like Qwen2.5-32B.

In contrast, smells related to user intent comprehension, such as \textit{Ambiguous Instruction} (0.67\%\allowbreak--2.20\%) and \textit{Incomplete Instruction} (0.31\%\allowbreak--3.30\%), exhibit consistently low frequencies across all models. 
This indicates that intent understanding is no longer the primary bottleneck in multi-turn interactions. 
Instead, the core challenges have shifted to maintaining contextual consistency: models struggle to implement modifications without violating invariants or introducing functional regressions. 
Furthermore, \textit{Code Rollback} rates are uniformly low (mostly below 4\%), indicating that models rarely revert to previously fixed erroneous states without justification.

\input{tab/rq2_LLM}

\begin{summarybox}
\textbf{RQ2 Answer:} {\smell} are prevalent across mainstream LLMs, with \textit{Must-Do Omit} and \textit{Partial Functionality Breakdown} identified as the most common defect types across all models.

\end{summarybox}

%% file: tab/rq2_LLM.tex
\definecolor{freshblue}{RGB}{0, 128, 128}

\newcommand{\gc}[2]{\cellcolor{freshblue!#1}#2\%}

\begin{table*}[t]
    \centering
    \caption{Distribution of Interaction Smells Across mainstream LLMs.}
    \label{tab:smell_heatmap}
    \resizebox{\textwidth}{!}{%
        \begin{tabular}{c|ccccccccc}
            \toprule
            \textbf{Model} & 
            \textbf{\makecell{Ambiguous\\Instr.}} & 
            \textbf{\makecell{Incomplete\\Instr.}} & 
            \textbf{\makecell{Must-Do\\Omit}} & 
            \textbf{\makecell{Must-Not\\Violate}} & 
            \textbf{\makecell{Signature\\Mismatch}} & 
            \textbf{\makecell{Incon-\\sistency}} & 
            \textbf{\makecell{Func.\\Breakdown}} & 
            \textbf{\makecell{Code\\Rollback}} & 
            \textbf{\makecell{Repetitive\\Resp.}} \\ 
            \midrule
            
            GPT-4o & 
            \gc{5}{0.67} & \gc{8}{1.67} & \gc{48}{58.86} & \gc{7}{1.34} & \gc{9}{2.01} & \gc{16}{7.02} & \gc{44}{34.78} & \gc{12}{3.68} & \gc{37}{48.83} \\
            
            DeepSeek-Chat & 
            \gc{9}{2.20} & \gc{11}{3.30} & \gc{44}{50.00} & \gc{7}{1.10} & \gc{11}{3.30} & \gc{10}{2.75} & \gc{36}{33.52} & \gc{8}{1.65} & \gc{33}{28.57} \\
            
            Gemini 2.5 Flash & 
            \gc{7}{1.08} & \gc{5}{0.54} & \gc{55}{78.65} & \gc{7}{1.08} & \gc{6}{0.81} & \gc{14}{5.14} & \gc{18}{8.65} & \gc{9}{2.16} & \gc{47}{56.76} \\
            
            Qwen2.5-32B & 
            \gc{6}{0.94} & \gc{4}{0.31} & \gc{49}{62.38} & \gc{10}{2.51} & \gc{12}{3.45} & \gc{16}{6.90} & \gc{46}{55.17} & \gc{11}{3.13} & \gc{39}{39.50} \\
            
            Qwen2.5-72B & 
            \gc{8}{1.45} & \gc{6}{0.97} & \gc{44}{50.24} & \gc{12}{3.38} & \gc{14}{4.83} & \gc{13}{4.35} & \gc{41}{44.44} & \gc{4}{0.48} & \gc{37}{34.30} \\
            
            Qwen3-235B-a22b & 
            \gc{6}{0.92} & \gc{10}{2.29} & \gc{46}{54.13} & \gc{15}{5.50} & \gc{12}{3.67} & \gc{15}{5.96} & \gc{38}{37.16} & \gc{12}{3.67} & \gc{36}{32.57} \\
            
            \bottomrule
        \end{tabular}%
    }
\end{table*}

%% file: sec/6_InCE.tex
\section{RQ3: Mitigation Strategy: The {\tool} Framework}
\label{mit-stra}
In this section, we present the design of {\tool}, a framework tailored to mitigate {\smell} by enforcing global constraints and conducting proactive quality checks.

\subsection{Experimental Design}
To address the prevalent issues of \textit{Must-Do Omit} and \textit{Partial Functionality Breakdown}, we propose Invariant-aware Constraint Evolution ({\tool}), a dedicated framework for explicit logical management. Rather than replacing the conventional full-history concatenation paradigm, {\tool} serves as an augmentation layer that decouples constraint handling from code generation. 
The core design motivation is to counteract attention dilution, a phenomenon where redundant and fragmented historical information obscures critical requirements as dialogue progresses. 
To achieve this, {\tool} adopts a decoupled architecture comprising two key modules: the Invariant Extraction Module (IEM), which distills global constraints, and the Proactive Smell Detector (PSD), which performs pre-generation quality auditing. This design ensures that high-priority logical invariants remain in focus within complex interaction histories, significantly enhancing the reliability of iterative code generation. 

The IEM leverages GPT-4o to identify and eliminate transient, localized, or superseded redundant instructions, retaining only global constraints that remain relevant across subsequent iterations. 
To resolve potential instructional conflicts, the IEM follows the Latest Instruction Priority principle, neutralizing interference from obsolete constraints. The core utility of this module lies in constructing a high-priority constraint list that preserves essential historical context. This list effectively mitigates attention dispersion in long-context interactions and prevents the model from regressing to prior erroneous states due to omitted key constraints.

The PSD serves as a quality auditor prior to task execution by the generator, ensuring the completeness and consistency of input instructions. It systematically detects potential {\smell}, including \textit{Ambiguous Instruction} and \textit{Cross-Turn Inconsistency}, through cross-referencing the current user intent with the invariant pool maintained by the IEM. Leveraging GPT-4o, it neutralizes semantic conflicts between new and historical instructions while automatically repairing logical inconsistencies. Ultimately, the PSD outputs a highly structured Constraint Checklist. This stage delivers clear and deterministic guidance, thereby precluding functional regressions arising from semantic misunderstandings or historical oversight. 

To better evaluate the effectiveness of the method, we introduced the Task Success Rate and Average Turns to Success metrics.

\textbf{Task Success Rate (TSR)}: serves as the primary metric for evaluating the end-to-end effectiveness of the model. It is defined as the percentage of tasks where the model successfully satisfies the user's intent within the maximum allowable interaction turns ($T_{max}=11$). A task is deemed successful if and only if the evaluation oracle assigns a score of 9 or higher to the generated code. Formally, TSR is calculated as:

\begin{equation}
  \text{TSR} = \frac{N_{success}}{N_{total}} \times 100\%
\end{equation}

where $N_{total}$ is the total number of evaluation tasks (170 per model), and $N_{success}$ is the count of tasks meeting the success criteria.

\textbf{Average Turns to Success (ATS)}: measures the interaction efficiency or the cost of communication required to solve a task. It computes the arithmetic mean of interaction turns consumed to reach the success state, calculated exclusively for the subset of successful tasks. ATS is defined as:

\begin{equation}
  \text{ATS} = \frac{\sum_{i \in S} \text{Turns}_i}{|S|}
\end{equation}

where $S$ represents the set of successful tasks, and $\text{Turns}_i$ denotes the number of rounds used in the $i$-th successful session. A lower ATS indicates that the model can resolve issues and converge to the correct solution with fewer rounds of refinement.

\subsection{Results}
\subsubsection{Overall Effectiveness}
Table ~\ref{tab:rq3_comparison} presents the overall performance comparison of the Invar-Evolution framework across six mainstream LLMs. 
Overall, the framework significantly enhances the models' capability to resolve complex programming tasks within multi-turn interactions, achieving improvements in Task TSR for 5 out of the 6 evaluated models. Notably, ChatGPT-4o and Gemini-2.5-Flash achieved the most substantial gains, with TSR increasing by 5.00\% and 6.67\%, respectively. These results confirm that explicit constraint management effectively mitigates the Must-Do Omit smell in multi-turn interactions. Even for models with robust baselines like DeepSeek-Chat and Qwen3-235B-a22b (TSR > 90\%), the framework facilitated further marginal gains, pushing both to a top-tier performance of 93.33\%. While Qwen2.5-72B exhibited a slight performance fluctuation (-3.34\%), this can be attributed to a ceiling effect resulting from its near-perfect baseline, with the performance still maintaining an SOTA level.

\input{tab/rq3_1}

Regarding interaction efficiency, the ATS shows a slight upward trend for most models (e.g., ChatGPT-4o increasing from 3.22 to 3.54). 
This increase does not indicate diminished efficiency. 
Instead, it arises from survivorship bias: InCE successfully resolves complex, long-context tasks that baseline methods fail to complete. 
These inherently demanding tasks necessitate more interaction steps; their inclusion in the successful sample set thus elevates the overall ATS average.
However, Gemini-2.5-Flash presents a compelling exception, achieving a ``dual victory'' where TSR increased while ATS decreased from 4.10 to 3.80. This suggests that for distraction-prone models, the framework accelerates convergence by eliminating futile \textit{Repetitive Response}. Overall, the marginal increase in ATS represents an acceptable computational cost for maintaining logical consistency, particularly when weighed against the significant gains in task success.

\subsubsection{Smell Mitigation}

To investigate the underlying causes of the observed performance gains, we conducted a granular analysis of how the InCE framework mitigates specific {\smell}. 
As shown in Table ~\ref{tab:smell_mitigation_final}, the framework consistently reduced the prevalence of critical smells across the three representative models: ChatGPT-4o, Gemini-2.5-Flash, and Qwen2.5-32B.
First, regarding the pervasive Must-Do Omit smell, a consistent decline was observed across the board. Beyond the substantial reduction in ChatGPT-4o (from 58.86\% to 45.83\%), even models with historically high omission rates, such as Gemini-2.5-Flash and Qwen2.5-32B, recorded reductions of around 8\% and 4\% respectively.
This improvement confirms that the IEM effectively preserves global constraints, preventing their omission even within extended contexts.
Secondly, when examining the aspect of Functional Breakdown, for Qwen2.5-32B—the model most severely affected by destructive updates—the prevalence decreased from 55.17\% to 50.12\%. Meanwhile, ChatGPT-4o also exhibited a remarkable improvement, with its relevant metric dropping from 34.78\% to 29.38\%. These results indicate that the PSD successfully acted as a critical safeguard against functional regression, significantly enhancing code stability.
Finally, in terms of \textit{Repetitive Response}, the framework showcased superior corrective capabilities, achieving double-digit improvements across models. The improvements extended beyond Qwen2.5‑32B (‑11.3\%) and Gemini‑2.5‑Flash (‑13.5\%); ChatGPT‑4o also registered a substantial decline, falling from 48.83\% to 36.79\%. 
These improvements directly validate the optimization logic of the ATS metric mentioned earlier—demonstrating that the framework can effectively identify and disrupt futile error loops, forcing models to break out of local optima and thereby accelerating task convergence.

\input{tab/rq3_smells}

\begin{summarybox}
\textbf{RQ3 Answer:} Our {\tool} can improve the TSR of most evaluation models. Crucially, the framework effectively mitigates {\smell}, particularly addressing subtypes such as \textit{Must-Do Omit} and \textit{Partial Functionality Breakdown}.

\end{summarybox}


%% file: tab/rq3_1.tex
\begin{table}[t]
    \centering
    \small 
    \caption{Performance Comparison: Vanilla Models vs. Ours.}
    \label{tab:rq3_comparison}
    
    \setlength{\tabcolsep}{6pt} 
    \begin{tabular}{lcccc}
        \toprule
        \multirow{2}{*}{\textbf{Model}} & \multicolumn{2}{c}{\textbf{TSR (\%)} $\uparrow$} & \multicolumn{2}{c}{\textbf{ATS} $\downarrow$} \\
        \cmidrule(lr){2-3} \cmidrule(lr){4-5}
        & Vanilla & \textbf{Ours} & Vanilla & \textbf{Ours} \\
        \midrule
        
        ChatGPT-4o & 
        78.33 & \textbf{83.33} & 
        \textbf{3.22} & 3.54 \\
        
        Gemini 2.5 Flash & 
        70.00 & \textbf{76.67} & 
        4.10 & \textbf{3.80} \\
        
        DeepSeek-Chat & 
        91.67 & \textbf{93.33} & 
        \textbf{2.31} & 2.43 \\
        
        Qwen2.5-32B & 
        76.67 & \textbf{80.00} & 
        \textbf{3.59} & 3.90 \\
        
        Qwen2.5-72B & 
        \textbf{96.67} & 93.33 & 
        3.19 & \textbf{3.11} \\
        
        Qwen3-235B-a22b & 
        91.67 & \textbf{93.33} & 
        \textbf{2.95} & 3.05 \\
        
        \bottomrule
    \end{tabular}
    
\end{table}


    
    
        
        
        
        
        
        
        

%% file: tab/rq3_smells.tex
\begin{table*}[t]
    \centering
    \caption{Mitigation of Interaction Smells.}
    \label{tab:smell_mitigation_final}
    
    \renewcommand{\arraystretch}{1.2} 
    \setlength{\tabcolsep}{3pt}       
    
    \resizebox{\textwidth}{!}{%
        \begin{tabular}{llccccccccc}
            \toprule
            \multirow{2}{*}{\textbf{Model}} & 
            \multirow{2}{*}{\textbf{Method}} & 
            \textbf{\makecell{Ambiguous\\Instr.}} & 
            \textbf{\makecell{Incomplete\\Instr.}} & 
            \textbf{\makecell{Must-Do\\Omit}} & 
            \textbf{\makecell{Must-Not\\Violate}} & 
            \textbf{\makecell{Signature\\Mismatch}} & 
            \textbf{\makecell{Cross-Turn\\Incon.}} & 
            \textbf{\makecell{Func.\\Breakdown}} & 
            \textbf{\makecell{Code\\Rollback}} & 
            \textbf{\makecell{Repetitive\\Resp.}} \\ 
            \midrule
            
            \multirow{2}{*}{GPT-4o} 
            & Vanilla & 0.67\% & 1.67\% & 58.86\% & 1.34\% & 2.01\% & 7.02\% & 34.78\% & 3.68\% & 48.83\% \\
            & \textbf{Ours} & 0.67\% & \textbf{1.00\%} & \textbf{45.83\%} & \textbf{0.67\%} & \textbf{1.67\%} & \textbf{2.01\%} & \textbf{29.38\%} & \textbf{2.34\%} & \textbf{36.79\%} \\
            \midrule
            
            \multirow{2}{*}{Gemini 2.5 Flash} 
            & Vanilla & 1.08\% & 0.54\% & 78.65\% & 1.08\% & 0.81\% & 5.14\% & 8.65\% & 2.16\% & 56.76\% \\
            & \textbf{Ours} & \textbf{0.54\%} & \textbf{0.27\%} & \textbf{70.58\%} & \textbf{0.54\%} & \textbf{0.27\%} & \textbf{1.60\%} & \textbf{6.56\%} & \textbf{1.62\%} & \textbf{43.24\%} \\
            \midrule
            
            \multirow{2}{*}{Qwen2.5-32B} 
            & Vanilla & 0.94\% & 0.31\% & 62.38\% & 2.51\% & 3.45\% & 6.90\% & 55.17\% & 3.13\% & 39.50\% \\
            & \textbf{Ours} & \textbf{0.63\%} & 0.31\% & \textbf{58.30\%} & \textbf{1.25\%} & \textbf{1.88\%} & \textbf{2.51\%} & \textbf{50.12\%} & \textbf{2.51\%} & \textbf{28.21\%} \\
            
            \bottomrule
        \end{tabular}%
    }
\end{table*}


%% file: sec/7_discussion.tex
\section{Discussion}
We provide practical design guidelines for future human-LLM interaction systems and discuss the threats to the validity of our study.
\subsection{Design Guidelines for Human–LLM Interaction Systems}

\noindent \textbf{Maintaining Explicit Constraint for Persistent Invariants}: A primary cause of interaction degradation is that critical constraints are embedded implicitly within long dialogue histories, making them easy for models to overlook as interactions evolve. 
To address this, future interaction systems should introduce an explicit constraint maintenance that extracts, summarizes, and persists invariant requirements separately from raw conversation text. Specifically, the system can extract explicit constraints such as formatting rules, functional requirements, or forbidden behaviors, and inject them into each generation step with high priority; LLMs can be guided to preserve previously satisfied constraints. This mechanism directly targets \textit{Must-Do Omission} and \textit{Must-Not Violate} smells by transforming implicit conversational memory into an explicit and enforceable state.

\noindent \textbf{Scoped Modification Authority to Control Model Actions}:  
Many functional regressions observed in multi-turn interactions stem from the model implicitly assuming global authority over the entire codebase when responding to localized refinement requests. 
Without an explicit notion of action scope, the model freely rewrites stable and previously validated code while attempting to satisfy new requirements, leading to destructive updates. 
To address this, interaction frameworks should explicitly grant LLMs scoped modification authority, clearly defining which regions of code the model is permitted to modify in a given turn and which regions must be treated as read-only context. By binding each interaction turn to an explicit action scope, such as ``modify only function X'' or ``update logic without changing interfaces'', the model’s generative behavior is aligned with user intent and constrained responsibility. This guideline reframes protection not as rigid locking, but as principled delegation of authority, closely mirroring change-scoping and ownership boundaries in software engineering. Explicit action scoping directly reduces \textit{Partial Functionality Breakdown} and \textit{Code Rollback} by preventing the model from performing unauthorized global rewrites during incremental evolution.

\noindent \textbf{Pre-Generation Smell Detection and Interaction Gating}: 
Several {\smell}, such as ambiguity, missing specifications, or conflicting constraints, can be identified before generation occurs. One can integrate a pre-generation smell detector to analyze new user instructions against existing constraints and historical decisions, blocking or redirecting generation when issues are detected. Instead of producing potentially faulty outputs, the system can request clarification or resolve conflicts proactively. This interaction gating mechanism prevents error propagation and reduces downstream regressions, particularly for \textit{Ambiguous Instruction}, \textit{Incomplete Instruction}, and \textit{Cross-Turn Inconsistency} smells.

\subsection{Threats to Validity}

\textbf{External Validity.}
Threats to external validity concern the generalizability of our findings. First, while we utilized large-scale real-world datasets such as WildChat and LMSYS-Chat-1M, filtered for coding-related content, they may not fully capture niche programming domains. Second, our evaluation involved six mainstream LLMs (e.g., GPT-4o, DeepSeek, Qwen) to represent the current SOTA. However, given the rapid evolution of model architectures, the applicability of our findings to future models remains to be verified.

\textbf{Internal Validity.}
The primary threat to internal validity stems from the subjectivity involved in constructing the interaction smell taxonomy and manually labeling data. To mitigate this, we employed a rigorous Open Card Sorting procedure. Two authors with expertise in software engineering independently coded the interaction logs, with a third senior researcher serving as an arbitrator to resolve disagreements. The inter-rater reliability, measured by Cohen’s Kappa, was 0.78, indicating a substantial level of agreement. Second, for the large-scale labeling and detection processes in RQ2 and RQ3, we leveraged the thinking mode of the Gemini-3-Flash-Preview model to assist in comprehending and annotating complex interaction contexts. Although we validated the accuracy of this LLM-assisted approach on a benchmark dataset, minor false positives or negatives may persist. To minimize this impact, we conducted a manual review of anomalous data points (outliers) during the experimental analysis to ensure the reliability of our conclusions.

\textbf{Construct Validity.}
Threats to construct validity relate to whether the metrics used accurately and comprehensively reflect interaction quality and collaboration efficiency. In the evaluation of RQ3, we exclusively utilized TSR and ATS to quantify the effectiveness of the InCE framework. However, we acknowledge the limitations of these objective metrics in gauging subjective user experience. Specifically, as objective, outcome-oriented metrics, TSR and ATS cannot capture the user's subjective sentiments during the collaboration process, such as satisfaction, frustration, or cognitive load, nor do they directly reflect the maintainability or stylistic preferences of the generated code. Nevertheless, to guarantee the objectivity and reproducibility of our evaluation and to avoid introducing unquantifiable subjective biases, this study focuses primarily on the functional outcome and convergence efficiency of the interactions. We leave the subjective assessment of user experience to future work.

%% file: sec/8_conclusion.tex
\section{Conclusion}
\label{sec:conclusion}
In this paper, we present the first systematic study on {\smell} in human-LLM collaborative programming. Through the construction of a comprehensive taxonomy and an empirical evaluation of six SOTA LLMs, we identify  \textit{Must-Do Omission} as the most prevalent obstacle to contextual consistency. To address this, we propose InCE, a multi-agent framework that safeguards interaction quality via explicit invariant extraction and pre-generation auditing. Our evaluations demonstrate that InCE significantly improves task success rates while effectively suppressing {\smell}, highlighting the critical need to prioritize process-oriented interaction quality in AI-assisted development.
